# Analogies between optical propagation and heat diffusion: Applications to micro-cavities, gratings and cloaks

C. Amra[1], D. Petiteau[1], M. Zerrad[1], S. Guenneau[1], G. Soriano[1], B. Gralak[1], M. Bellieud[2], D. Veynante[3], N. Rolland[4]

[1]Aix Marseille Université, Institut Fresnel, CNRS, Ecole Centrale Marseille
Faculté des Sciences et Techniques de St Jérôme, 13397 Marseille Cedex 20
[2]Université Montpellier 2, CNRS, LMGC
[3]Ecole Centrale Paris, CNRS, EM2C
[4]Université de Lille 1, CNRS, IEMN
claude.amra@fresnel.fr

**Abstract :**
A new analogy between optical propagation and heat diffusion in heterogeneous anisotropic media has been proposed recently [S. Guenneau, C. Amra, and D. Veynante, Optics Express Vol. **20**, 8207-8218 (2012)]. A detailed derivation of this unconventional correspondence is presented and developed. In time harmonic regime, all thermal parameters are related to optical ones in artificial metallic media, thus making possible to use numerical codes developed for optics. Then the optical admittance formalism is extended to heat conduction in multilayered structures. The concepts of planar micro-cavities, diffraction gratings, and planar transformation optics for heat conduction are addressed. Results and limitations of the analogy are emphasized.

## 1- Introduction

Optics and heat conduction have often been associated, since optical devices and systems generally have to face thermal effects. When the electromagnetic field interacts with dense matter, optical losses are converted into heat, which generates conductive, convective, and radiative transfer. These well-known photo-induced thermal effects have motivated a number of papers, for instance in the field of photo-thermal microscopy, laser-induced damage, and temperature insensitive devices [1-6]. Specific models have then been developed to take into account heat transfer inside multilayers [1-6]. Moreover, thermal emission has been deeply investigated, within the framework of nanoscale photonic structures [7-15].

A novel association of optics and heat has been proposed recently [16-29]. It is based on an analogy between the governing equations of propagation for optics and diffusion for heat. The correspondence is not intuitive according to the markedly different nature of optical propagation and thermal diffusion phenomena. Nevertheless, an analogy can be unveiled in time harmonic regime when metallic media are considered in optics. Under these conditions, electromagnetic fields decay exponentially fast away from a metal-dielectric interface so that their behaviour is somewhat reminiscent of a temperature field undergoing a fast decay away from a heat source. Such an analogy has been first used in the design of thermal cloaks, heat concentrators and rotators [16-29], using geometric transformation methods previously applied to electromagnetics [30-31]. A series of papers subsequently emerged and confirmed tools of transformation optics can be applied to control other diffusion processes [32-33].

In the present work, we further explore the unconventional analogy between transformation optics and thermodynamics which has been first proposed in [16]. Correspondences are emphasized between optical and thermal fields, i.e. (vector) electric fields and (scalar) temperature, (vector) magnetic fields and (vector) heat flux. All thermal parameters are related to optical parameters in artificial metallic media, thus making it possible to reintroduce the optical concepts of effective index and complex admittance for heat conduction. More generally, all numerical codes developed for optics (multilayers, gratings, micro-cavities, scattering and diffraction, transformation optics…) can be used. The plan of this article is as follows: the whole optical admittance formalism [34-39] is first extended to thermal conduction, and deep analogies are drawn with multilayered electromagnetic structures in order to predict thermal properties in multilayers. Next, the concepts of planar micro-cavities, diffraction gratings and planar cloaks are addressed in the frame of heat conduction. A particular attention is paid to the mathematical correspondences and limitations between metal optics and heat conduction, including fields and energy balances, together with the extended optical admittance formalism for conduction.

## 2- Optical propagation versus heat conduction

### 2-a Spatio-temporal regime

Free space optical propagation in linear, isotropic and homogeneous materials can be classically modelled using the governing equation [39-44]:

$$\Delta \mathbf{E} - \varepsilon\mu *_t \partial^2 \mathbf{E}/\partial t^2 = S_{opt}(\mathbf{J},q) \quad (1)$$

with $\mathbf{E}$ the vector electric field, $\varepsilon$ and $\mu$ the temporal permittivity and permeability, $\mathbf{J}$ and $q$ the density of currents and charges and $S_{opt}$ a source term resulting from these densities. In equation (1) the convolution product ($*_t$) involves time (t) and takes into account the material inertia, that is, the time delay between the excitation of material and its optical response. Note the second-order derivation versus time of the electric field (elliptic equation).

On the other hand, heat conduction [44-46] involves a first-order derivation versus time (parabolic equation). We will here consider that conduction satisfies the following equation in linear, isotropic and homogeneous media as:

$$\Delta T - 1/a \, \partial T/\partial t = -S/b = S_{th} \quad (2)$$

with T the temperature and a,b the thermal parameters, that is, diffusivity and conductivity, respectively. The source term S is the bulk density of heat power. Note that in equation (2) the possible temperature dependence of the thermal parameters is neglected.

In addition to the discrepancy in time derivation order, other straightforward differences can be emphasized between optical propagation and heat diffusion, which are:
- Heat equation is scalar, while optics equation are vectorial; however one can consider optics equations on each basis vector
- Optical parameters ($\varepsilon,\mu$) vary with time and hence will exhibit a frequency dispersion, while this is not the case for the thermal parameters in equation (2)

Another difference lies in the fact that the temperature (or a flux condition) is most often fixed at the domain frontier, a supplementary condition not present in free space optics. However, in the absence of sources heat flux and temperature are continuous at interfaces, as it is for the tangential electromagnetic fields.

In equations (1-2) all fields vary with time (t) and space location $\boldsymbol{\rho} = (\mathbf{r},z) = (x,y,z)$. It is interesting to recall the Green's functions G of equations (1-2) that give the particular solutions in the form $G *_t S_{opt}$ or $G *_t S_{th}$, with:

$$G_{opt}(\boldsymbol{\rho},t) = (-1/4\pi\rho) \, \delta(t-\rho\sqrt{(\varepsilon\mu)}0.5) \quad \text{in optics} \quad (3)$$
$$G_{th} = (1/8b\sqrt{a}) \, [1/(\pi t)^{3/2}] \exp[-\rho^2/(4at)] \, H(t) \quad \text{in conduction} \quad (4)$$

with $\rho = |\boldsymbol{\rho}|$, H the Heaviside function and $\delta$ the Dirac distribution. Equation (3) is given under the assumption that the optical material is not dispersive, and shows that the optical wave-front propagates at velocity $v = 1/\sqrt{(\varepsilon\mu)}$ and is located at time t at the sphere surface of radius $\rho = vt$ where the whole energy is located. On the other hand, equation (4) is for a diffusion process, so that the energy is spreaded within the whole heated volume of radius $\rho = 2\sqrt{(at)}$ which increases with time. Notice that (4) does not involve any Dirac distribution: strictly speaking heat would have already diffused everywhere at arbitrary time ($t\neq 0 \Rightarrow G_{th} \neq 0 \, \forall \, \rho$), due to the Gaussian nature of $G_{th}$.

In what follows we now consider the propagation or conduction equations in isotropic homogeneous regions free of sources, what we write as the homogeneous equations:

$$\Delta E - \varepsilon\mu *_t \partial^2 E/\partial t^2 = 0 \qquad (5) \qquad \text{for optics, on each basis vector } (E = E_x, E_y \text{ or } E_z)$$
$$\Delta T - 1/a\; \partial T/\partial t = 0 \qquad (6) \qquad \text{for conduction}$$

Moreover, we will be interested in a temperature elevation resulting from a transient heat source, similar to what happens when the heat source results from the absorption of a modulated laser beam. Following the superposition properties of linear systems, the total temperature satisfies $T = T_a + T'$ with $T_a$ due to the environment, and $T'$ resulting from the heat source given as $S = S_0 [1 + h(t)]$ with $|h(t)| < 1$. This last temperature $T'$ can be written as $T' = T_1 + T_2$, where $T_1$ is a steady-state temperature and results from $S_0$, while $T_2$ is a transient temperature and results from $S_0 h(t)$. In this work we are actually interested in this last transient temperature $T_2$.

## 2-b Harmonic regime

For the sake of simplicity we use Fourier transform rather than Laplace transform, despite the differences in the two transforms. Another reason is that a modulated laser beam will create optical absorption (and so heat power) resulting in a periodic h(t) function which can be written as a Fourier series. Hence we now consider Fourier transform versus time of the optical and heat equations, with ω the temporal pulsation, conjugate of the time variable, and $f = \omega/2\pi$ the temporal frequency. The result is well-known and written as:

$$\Delta \breve{u} + k^2(\omega)\, \breve{u} = 0 \qquad (7)$$

with k the wavenumber (or conduction number- by analogy with optics), u the scalar electric field or temperature, and $\breve{u}$ its Fourier transform defined as:

$$\breve{u}(\boldsymbol{\rho},f) = \int_t u(\boldsymbol{\rho},t)\exp(j2\pi ft)\,dt \qquad (8\text{-}a) \qquad \text{and} \qquad u(\boldsymbol{\rho},t) = \int_f \breve{u}(\boldsymbol{\rho},f)\exp(-j2\pi ft)\,df \qquad (8\text{-}b)$$

Equation (7) shows that both fields (u = E or T) satisfy a similar harmonic equation in the Fourier space, taking into account the specific dispersion law of the k function, that is:

$$k^2(\omega) = \omega^2 \breve{\varepsilon}(\omega)\breve{\mu}(\omega) \qquad \Leftrightarrow \qquad k = \omega\sqrt{[\breve{\varepsilon}(\omega)\breve{\mu}(\omega)]} \qquad \text{in optics} \qquad (9)$$
$$k^2(\omega) = j\omega/a \qquad \Leftrightarrow \qquad k = (1+j)\sqrt{[\omega/2a]} \qquad \text{in conduction} \qquad (10)$$

In other words, the dispersion law k(ω) is the memory of the pth order of time derivation (p = 1 for conduction, p = 2 for optics). Notice in (9) that $\breve{\varepsilon}(\omega)$ and $\breve{\mu}(\omega)$ are the Fourier transforms of the optical functions (ε,μ), while in (10) the diffusivity parameter a is not dispersive under the thermal approximation. Hence the k dispersion behaves as √ω for conduction, while in optics it depends on the permeability and permittivity frequency dispersion law; however as a first approximation the k dispersion in optics can be considered to follow ω in a narrow bandwidth.

Above all a major point to be emphasized in the harmonic regime concerns the k value at a given frequency. Indeed in optics such a value can be real or complex, depending upon whether dielectrics or metals are under study. On the other hand, the k value is necessarily complex in conduction for reasons following (7-10), so that heat diffusion can be considered to be analogous to optical propagation (more exactly, electric field attenuation) in artificial metallic media. Furthermore, following (10) the equivalent metal should be specific with identical real and imaginary parts in the refractive index n (defined as $k = k_0 n$ in optics).

## 3- Optical admittance formalism for multilayers

### 3-a Multilayer Optics: basics

In multilayer optics (figure 1) the admittance formalism is commonly used to predict optical properties and to design antireflective coatings, beam splitters, polarizing devices, mirrors and narrow-band filters [34-39]... It is based on the complex admittance, a key function that often simplifies the calculation and allows for analytical design. This admittance function Y is given for a particular polarization (transverse electric S or transverse magnetic P) as:

$$\widehat{\boldsymbol{H}} = Y\, \mathbf{z} \wedge \widehat{\boldsymbol{E}} \qquad (11)$$

with $\widehat{\boldsymbol{H}}$ and $\widehat{\boldsymbol{E}}$ the tangential components of the electric and magnetic polarized field, and **z** the direction normal to the stack. Notice here that all fields are harmonic (monochromatic) and result from a single illumination incidence at the top surface of the stack; in other words, the admittance is defined after double Fourier transform (versus time t and versus transverse coordinate **r**) of the fields; we will come back to this point in the next section where we re-derive the admittance formalism.

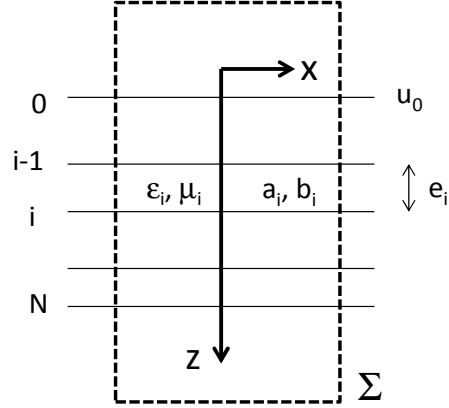

*Figure 1: Geometry of the multilayer for optics or heat conduction*

In the case of a stationary wave within the stack, the admittance varies with the z altitude; on the other hand, when the wave is only progressive or retrograde, what may happen in free space in the surrounding media, it is constant and reduced to the effective index ±m, that is:

$$\hat{H}^+ = m\, \mathbf{z} \wedge \hat{E}^+ \quad \text{for a progressive wave} \quad (12\text{-a})$$
$$\hat{H}^- = -m\, \mathbf{z} \wedge \hat{E}^- \quad \text{for a retrograde wave} \quad (12\text{-b})$$

Equations (11-12) are given for each polarization of the optical field, which means that both effective indices and admittances are polarization (and incidence, wavelength…) dependent. Moreover we have assumed to be in the far field, so that the electric field **E** is driven by 2 transverse polarization modes.

The effective indices are known from the illumination conditions, and enable to initiate a sequence for the admittance extraction throughout the whole stack (see next section). They are given as:

$$m = (1/\eta_0 \mu_r)\, n\alpha/k \quad \text{for transverse electric (TE) polarization} \quad (13\text{-a})$$
$$m = (1/\eta_0 \mu_r)\, nk/\alpha \quad \text{for transverse magnetic (TM) polarization} \quad (13\text{-b})$$

with $\eta_0 = \sqrt{(\mu_0/\varepsilon_0)}$ the vacuum impedance, $n = \sqrt{(\check{\varepsilon}_r \check{\mu}_r)}$ the refractive index, $(\check{\varepsilon}_r, \check{\mu}_r)$ the complex relative permittivity and permeability, $\alpha = \sqrt{(k^2 - \sigma^2)}$ and $\sigma$ the spatial pulsation which can be connected in transparent media to a propagation direction $\theta$ (illumination angle), that is:

$$\sigma = k\sin\theta = 2\pi\nu \quad \Rightarrow \quad \alpha = k\cos\theta \quad (14)$$

with $\nu$ the spatial frequency; in the case of plane waves ($\sigma < k$), relation (13) is turned into:

$$m = (1/\eta_0 \mu_r)\, n\cos\theta \text{ for TE polarization} \quad \text{and} \quad m = (1/\eta_0 \mu_r)\, n/\cos\theta \text{ for TM polarization}$$

All relations (11-14) are valid for plane, evanescent or dissociated waves.

### 3-b Conduction in multilayers

Let us now derive the admittance formalism in a few lines, and show how it can also be applied to conduction. For the sake of simplicity (ie: to avoid the double Fourier transform addressed in the next sections), we start with a problem invariant with respect to the transverse coordinates (x,y), so that relation (7) only involves z and $\omega$ variables:

$$\partial^2 \breve{u}/\partial z^2 + k^2(\omega)\breve{u} = 0 \quad (15)$$

In optics the unique z-dependence corresponds to the case of a plane wave illuminating the stack at normal incidence ($\sigma = 0$). The solution of (15) is given in each layer (i=1, N) by:

$$\breve{u}_i(z,\omega) = \breve{u}_i^+(\omega) \exp[jk_i(\omega)z] + \breve{u}_i^-(\omega) \exp[-jk_i(\omega)z] \quad (16)$$

where the field is E or T in the layer. The next step in optics to complete the admittance formalism is based on the fact that the magnetic field $\breve{v}$ follows the same analytical expression (16), and that the electric and magnetic components are linked though (11-12), that is:

$$\breve{v}_i(z,\omega) = \breve{v}_i^+(\omega) \exp[jk_i(\omega)z] + \breve{v}_i^-(\omega) \exp[-jk_i(\omega)z] \quad (17)$$
$$\breve{v}_i^+ = m_i\, \mathbf{z} \wedge \breve{u}_i^+ \quad \text{and} \quad \breve{v}_i^- = -m_i\, \mathbf{z} \wedge \breve{u}_i^- \quad (18)$$
$$\breve{v}_i = \breve{v}_i^+ + \breve{v}_i^- = Y_i\, \mathbf{z} \wedge \breve{u}_i \quad (19)$$

The last step consists in combining (16-19) and writing the continuity of the tangential electric and magnetic fields in the absence of sources, which results in a matrix formalism to pass from one interface to the next as below:

$$(\mathbf{z}\wedge\breve{u}_{i-1}, \breve{v}_{i-1}) = M_i\, (\mathbf{z}\wedge\breve{u}_i, \breve{v}_i) \quad \text{with the matrix} \quad M_i = (s^i_{kl}) \quad (20)$$
$$s^i_{11} = s^i_{22} = \cos\delta_i,\ s^i_{12} = -j\sin\delta_i/m_i,\ s^i_{21} = -j\, m_i \sin\delta_i \ \text{and}\ \delta_i = k_i e_i \quad (21)$$

with $e_i$ the thickness of layer i.

Now we can check that the similarity with the conduction process is immediate. First of all, we need two quantities that are continuous throughout the whole stack; in the absence of singular sources, these quantities are the harmonic temperature $\breve{T}$ and the scalar (z-projection) heat flux density $\breve{h}$ given by:

$$\breve{h} = -b\, \partial\breve{T}/\partial z \quad \text{with} \quad \partial^2\breve{h}/\partial z^2 + k^2(\omega)\, \breve{h} = 0 \quad (22)$$

Hence the temperature plays the role of the electric field in (16), while the heat flux replaces the magnetic field in (17). However this last condition must be further clarified to determine which quantity replaces the effective index m in (18). By analogy with optics (relation 12), this last quantity is defined for conduction as:

$$\breve{h}^+ = m\breve{T}^+ \quad \text{and} \quad \breve{h}^- = -m\breve{T}^- \quad (23)$$

The result is immediate from (17), that is:
$$m = -jkb \quad (24)$$
and the admittance follows within the stack as:
$$\breve{h}(\omega,z) = \breve{h}^+ + \breve{h}^- = Y(\omega,z)\, \breve{T}(\omega,z) \quad (25)$$

At this stage the correspondence is complete and the matrix formalism (20-21) can be used either in optics or conduction in the harmonic regime. Notice that two slight modifications were necessary to reach this one-to-one correspondence, which are:
- the optical parameter $k_i = \omega\sqrt{[\breve{\varepsilon}_i(\omega)\breve{\mu}_i(\omega)]}$ is replaced by the conduction parameter $k_i = (1+j)\sqrt{[\omega/2a_i]}$
- the optical effective index $m_i = n_i/(\eta_0\mu_{ri})$ is replaced by that of conduction $m_i = -jk_i b_i$

## 3-c Wavelength, diffusion length and phase velocity

From equations (8) and (16) the field (T or E) can be reconstructed in the layers as the sum of elementary components. In the substrate these components take the form:

$$\breve{u}_{el} = \text{Real}\{\breve{u}^+(\omega) \exp[j(k(\omega)z-\omega t)]\} = \exp(-k''z)\, |\breve{u}^+|\, \cos[k'z - \omega t + \psi^+] \quad (26)$$

with $\breve{u}^+ = |\breve{u}^+|\exp(j\psi^+)$ and $k = k' + j k''$. The real exponential is for the field envelope while the cosine term allows one to retrieve a spatial period $\lambda = 2\pi/k'$ (the wavelength in optics) and a phase velocity $v = \omega/k'$ (light velocity in optics). Equation (26) shows that in conduction the (pseudo) spatial period $\lambda_{th}$ is proportional to the diffusion length

$$L = \sqrt{(2a/\omega)} \quad (27)$$

defined at 1/e for the temperature amplitude, that is: $\lambda_{th} = 2\pi L$. Also, quantity

$$v_{th} = \sqrt{(2a\omega)} \quad (28)$$

is analogous to the phase velocity of optics. However we do not extend the analogy to the refractive index, since this index involves the vacuum property in optics (where propagation occurs), while conduction fails in vacuum (another reference material should be used).

## 3-d Two-dimensional formalism for spatial wave packet

Now we take into account the spatial distribution of the field, which means that we drop the unique z-dependence to recover a $\boldsymbol{\rho} = (\mathbf{r},z)$ variable, with $\mathbf{r} = (x,y)$. Such an extension to a 3D geometry is immediate if we consider, in addition to the frequency wave packet (Fourier Transform versus time), a spatial wave packet (Fourier transform versus space coordinate $\mathbf{r}$). This second Fourier transform is given as:

$$\hat{u}(\boldsymbol{\nu},z,\omega) = \int_\mathbf{r} \breve{u}(\mathbf{r},z,\omega) \exp(-j2\pi\boldsymbol{\nu}\cdot\mathbf{r}) \, d\mathbf{r} \quad (29\text{-a}) \quad \text{and} \quad \breve{u}(\mathbf{r},z,\omega) = \int_{\boldsymbol{\nu}} \hat{u}(\boldsymbol{\nu},z,\omega) \exp(j2\pi\boldsymbol{\nu}\cdot\mathbf{r}) \, d\boldsymbol{\nu} \quad (29\text{-b})$$

with $\boldsymbol{\nu}$ the spatial frequency, the Fourier variable conjugated of $\mathbf{r}$. Such double Fourier transform $\hat{u}$ is governed by the same equation (15) as the single Fourier transform $\breve{u}$, except that $k(\omega)$ must be replaced by $\alpha(\omega) = \sqrt{k^2 - \sigma^2}$, with $\boldsymbol{\sigma} = 2\pi\boldsymbol{\nu}$ the spatial pulsation. We obtain:

$$\partial^2 \hat{u}/\partial z^2 + \alpha^2(\sigma,\omega)\hat{u} = 0 \quad (30)$$

In each layer the field in this second Fourier plane takes the form:

$$\hat{u}_i(\boldsymbol{\nu},z,\omega) = \hat{u}_i^+(\boldsymbol{\nu},\omega) \exp[j\alpha_i(\nu,\omega)z] + \hat{u}_i^-(\boldsymbol{\nu},\omega) \exp[-j\alpha_i(\nu,\omega)z] \quad (31)$$

so that the elementary components that enable to rebuild the original (spatio-temporal) field can once again be written in the substrate as:

$$\hat{u}_{el} = \text{Real}\{\hat{u}^+(\boldsymbol{\sigma}, \omega) \exp\{j[\boldsymbol{\sigma}\cdot\mathbf{r} + \alpha(\sigma,\omega)z - \omega t]\} = \exp(-\alpha''z) \, |\hat{u}^+| \cos[\boldsymbol{\sigma}\cdot\mathbf{r} + \alpha'z - \omega t + \phi^+] \quad (32)$$

with $\hat{u}^+ = |\hat{u}^+| \exp(j\phi^+)$ and $\alpha = \alpha' + j\alpha''$. In other words, this works like optics at oblique incidence and the k parameter only has to be replaced by $\alpha = \sqrt{k^2 - \sigma^2}$. The salient consequence for the matrix formalism (20-21) following the previous section is that the effective index for heat conduction (relation (24)) must also be modified as:

$$m = -j\alpha b \quad (33)$$

Regarding the spatial and temporal period, k' must be replaced by $\text{Real}(\alpha)$. At last the original (real) field is reconstructed with a double inverse Fourier Transform over temporal (f) and spatial ($\boldsymbol{\nu}$) frequency.

## 4- Energy balance

To be complete, our correspondences should also involve an energy balance for both optical propagation and heat conduction. This condition is written below within a domain $\Omega$ limited by a closed surface $\Sigma$ with unit outward normal $\mathbf{n}$ (see figure 1).

### 4-a Temporal regime

In optics we obtain from Maxwell's equations in the spatio-temporal regime:

$$-\int_\Omega \mathbf{J}\cdot\mathbf{E} \, dv = \int_\Sigma \mathbf{P}\cdot\mathbf{n} \, d\Sigma + \partial/\partial t \left(\int_\Omega w \, d\Omega\right) \Leftrightarrow F = \Phi + A \quad (34)$$

with $\mathbf{P} = \mathbf{E} \wedge \mathbf{H}$ the real Poynting vector, F the optical power provided by sources within $\Omega$, $\Phi$ the Poynting flux through the surface $\Sigma$, and A the variation of internal energy (or absorption) within $\Omega$, with $dA/d\Omega = \partial w/\partial t = \mathbf{H}\cdot\partial/\partial t(\mu*\mathbf{H}) + \mathbf{E}\cdot\partial/\partial t(\varepsilon*\mathbf{E})$. As usual the density of electromagnetic energy reduces to $w = (1/2)[\mathbf{E}^2 + \mathbf{H}^2]$ when the material inertia is neglected.

For purely heat conduction the balance is directly obtained from the governing equation integrated over $\Omega$:

$$S = \text{div}(\mathbf{h}) + (b/a) \, \partial T/\partial t \quad \text{with} \quad \mathbf{h} = -b \, \mathbf{grad}T \quad (35\text{-a})$$

$$\Rightarrow \int_\Omega S \, d\Omega = -\int_\Sigma b \, \mathbf{grad}T\cdot\mathbf{n} \, d\Sigma + \partial/\partial t \left[\int_\Omega (b/a) T \, d\Omega\right] \Leftrightarrow F = \Phi + A \quad (35\text{-b})$$

4with F the heat power provided by the source, Φ the heat flux through Σ and A the variation of internal energy, with: (b/a) $\partial T/\partial t = \partial w/\partial t = dA/d\Omega$. In other words, the density of electromagnetic energy is here replaced by a quantity proportional to the temperature, with b/a = γ C the product of mass bulk density (γ) and heat capacity (C).

## 4-b Fourier analysis

Such balances (34-35) must now be rewritten in the Fourier planes. In optics quadratic (intensity) detectors are known to realize a time average process of the square fields, due to their low temporal response with respect to optical periods. Such specificity allows one to transform the spatio-temporal balance (34) into a single harmonic balance at each temporal pulsation ω (first Fourier plane), that is:

$$-\int_\Omega \boldsymbol{J}.\boldsymbol{\breve{E}}^* \, d\Omega = 2 \int_\Sigma \boldsymbol{\Pi}.\boldsymbol{n} \, d\Sigma - j\omega\{\int_\Omega [\breve{\varepsilon}|\boldsymbol{\breve{E}}|^2 + \breve{\mu}|\boldsymbol{\breve{H}}|^2] d\Omega\} \tag{36}$$

with **Π** the complex harmonic Poynting vector: $\quad \boldsymbol{\Pi} = (1/2) \boldsymbol{\breve{E}}^* \wedge \boldsymbol{\breve{H}}$ (37)

Then to reach a balance at a single spatial frequency **ν** in the second Fourier plane, one may use Parseval's theorem and other specific properties to find:

$$-\int_{\boldsymbol{\nu},z} \boldsymbol{\hat{J}}.\boldsymbol{\hat{E}}^* d\boldsymbol{\nu} dz = 2 \int_{\boldsymbol{\nu}} \boldsymbol{\Pi}'.\boldsymbol{n} \, d\boldsymbol{\nu} - j\omega \{\int_{\boldsymbol{\nu},z} [\breve{\varepsilon}|\boldsymbol{\hat{E}}|^2 + \breve{\mu}|\boldsymbol{\hat{H}}|^2] \, d\boldsymbol{\nu} \, dz\} \tag{38}$$

$$\boldsymbol{\Pi}' = (1/2) \boldsymbol{\hat{E}}^* \wedge \boldsymbol{\hat{H}} \tag{39}$$

In a last step, considering the hermiticity of the fields we recover the harmonic balance per unit of spatial frequency $d\nu_x d\nu_y$ with respect to a surface Σ surrounding the whole planar multilayer with sides at infinity (figure 1):

$$-\text{Real}\{\int_z \boldsymbol{\hat{J}}.\boldsymbol{\hat{E}}^* dz\} = \text{Real}\{[Y|\boldsymbol{\hat{E}}|^2]_z\} + \omega\{\int_z [\varepsilon''|\boldsymbol{\hat{E}}|^2 + \mu''|\boldsymbol{\hat{H}}|^2] dz\} \tag{40}$$

with $\breve{\varepsilon}'' = \text{Im}(\breve{\varepsilon})$, $\breve{\mu}'' = \text{Im}(\breve{\mu})$ and the notation $[u]_z = u(z_2) - u(z_1)$, where $z_i$ are the upper and lower altitudes of Σ and $z_2 > z_1$. Notice here that we used equations (11) and (39) to reach the property: $2\boldsymbol{\Pi}'.\boldsymbol{n} = Y|\boldsymbol{\hat{E}}|^2$, a flux quantity which is continuous throughout the stack in the absence of sources.

Addressing the harmonic balance in conduction is a priori simpler since temporal frequencies are much lower than in optics, thus disregarding temporal integration by the detector is legitimate. Moreover, here the thermal parameters are assumed to be non dispersive, what simplifies the calculation. One can directly use the Fourier transform versus time of the governing equation (35) to reach a complex harmonic balance at temporal pulsation ω:

$$\int_\Omega \breve{S} d\Omega = -\int_\Sigma b \, \partial \breve{T}/\partial n \, d\Sigma - j\omega \int_\Omega (b/a) \breve{T} d\Omega \tag{41}$$

Per unit of surface area dxdy parallel to the interfaces of the stack, we obtain:

$$\int_z \breve{S} dz = -[b \, \partial \breve{T}/\partial z]_z - j\omega \int_z (b/a) \breve{T} dz \tag{42}$$

Due to the hermitian property of $\breve{T}$, such a balance can also be limited to positive frequencies ω as follows:

$$\text{Real}\{\int_z \breve{S} dz\} = -\text{Real}\{[b \, \partial \breve{T}/\partial z]_z\} - \text{Real}\{j\omega \int_z (b/a) \breve{T} dz\} \tag{43}$$

Now to write the conduction balance at a single spatial frequency (second Fourier plane), a z integration after a double Fourier transform of (35) directly leads to:

$$\int_z \hat{S} \, dz = [Y \hat{T}]_z - \int_z b \, \alpha^2 \hat{T} dz \tag{44}$$

with $\alpha^2 = k^2 - \sigma^2$. Notice again in (44) that $Y \hat{T}$ is a flux quantity which is continuous within the stack.

## 5- Optical admittance calculation for heat conduction

This section is devoted to numerical calculation. For that we use an optical thin film admittance software where the k dispersion and the effective index were rewritten following (10) and (33). As usual [34-38], the effective index value in the substrate is the starting value of the admittance sequence within the stack, given at each interface by:

$$Y_{i-1} = (Y_i \cos\delta_i - j \, m_i \sin\delta_i)/(\cos\delta_i - j \, Y_i \sin\delta_i/m_i) \tag{45}$$

A similar sequence is given for the field at interfaces:

$$\hat{u}_{i-1} = \hat{u}_i(\cos\delta_i - j \, Y_i \sin\delta_i/m_i) \tag{46}$$

A complete z-variation of admittance and field can also be obtained in the layers if we replace $\delta_i = \alpha_i e_i$ by $\delta_i = \alpha_i z$, with $0 < z < e_i$. Finally reflection (r) and transmission (t) factors can be defined in a way similar to optics, that is:

$$r = (m_0 - Y_0)/(m_0 + Y_0) \quad \text{at the top interface} \quad (47)$$
$$t = (1+r)/[\Pi_{i=1,N} (\cos\delta_i - j\, Y_i \sin\delta_i / m_i)] \quad \text{at the bottom interface} \quad (48)$$
$$r = \hat{u}_0^- / \hat{u}_0^+ \quad \text{and} \quad t = \hat{u}_{N+1}^+ / \hat{u}_0^+ \quad (49)$$

In optics the flux balance related to a far field surface surrounding the whole multilayer leads to:
$$\Phi_0^+ = \Phi_0^- + \Phi_S^+ + \Phi_0^+ A \quad (50)$$

with $\Phi$ the Poynting fluxes which are incident ($\Phi_0^+$), reflected ($\Phi_0^-$) and transmitted ($\Phi_S^+$), and A the normalized absorption. For an elementary plane wave these fluxes are proportional to $\Phi = \text{Real}(m)|\tilde{u}^\pm|^2$, so that the flux balance can be rewritten for the surface $\Sigma$ as:
$$1 = R + T + A, \text{ with } R = |r|^2 \text{ and } T = [\text{Real}(m_s)/\text{Real}(m_0)]|t|^2$$

The heat balance is slightly different in the sense that, due to heat diffusion, we cannot consider a source located at infinity ($z = -\infty$) so that the far field position of the source is not valid and may underpin the separation between incident and reflected fluxes. However the top interface temperature $T_0$ is assumed to be forced (and known), so that the whole field distribution can be obtained within the stack, from the sequence (46). Then the spectral quantity which we will analyze here is the ratio $t_{ht} = \hat{T}_S^+/\hat{T}_0 = t/(1+r)$, with $\hat{T}_s^+$ the field at the substrate interface. Hence in the figures of this section, all fields are normalized to $\hat{T}_0$.

Now some elementary or preliminary remarks are useful before numerical calculations. Indeed numerous tools were developed in optics to design optical coatings and address specific challenges, which can be briefly addressed here in conjunction with heat conduction.

## 5-a Preliminary remarks

### Single boundary
Consider first the case of a single boundary. As in optics, reflection is given by:
$$r = (m_0 - m_s)/(m_0 + m_s) = (-j\alpha_0 b_0 + j\alpha_s b_s)/(-j\alpha_0 b_0 - j\alpha_s b_s) \quad (51)$$

For a unique z-dependence (analogous to normal incidence) we obtain:
$$\alpha_i = k_i = (1+j)\sqrt{(\omega/2a_i)} \Rightarrow r = (\beta_0 - \beta_1)/(\beta_0 + \beta_1) \quad (52)$$
with $\beta_i$ the effusivity:
$$\beta_i = b_i/\sqrt{a_i} \quad (53)$$

This last equation shows that, when the calculation of reflection is considered, the effusivities in heat conduction play the role of the refractive indices in optics. This is valid at normal incidence, that is, for a zero spatial frequency ($\sigma = 0$). It is interesting to notice that the range of effusivities is much larger in conduction (> 2 decades) than in optics (half a decade). Besides from that, temperature reflection from a single boundary is a real quantity. We also obtain: $1 + r = t \Rightarrow t_{th} = 1$

### Specific multilayers
Absentee (half-wave) layers and quarter-wave stacks play a key role in multilayer optics [34-39], due to their stationary properties. For these stacks the phase term $\delta$ in the matrix (21) must be real and a multiple of $\pi/2$, which strongly simplifies the matrix and allows for an analytical design. However this is only valid for quasi-transparent materials ($\delta$ is a real number), while the analogy between optics and conduction is for metallic layers. Therefore the concept of quarter-wave stacks does not hold for conduction.

We could also wonder whether other concepts in metal optics could still be used for conduction. For instance induced transmission filters [34-38] combine dielectric and metallic films to produce narrow-band filters with large rejection bands; with such devices, the optical field can be confined within the spacer (cavity) layer in a way similar to all-dielectric Fabry-Perot filters. Unfortunately this will not be allowed for conduction multilayers: indeed induced transmission filters require mixing metallic and dielectric layers, while for the optics/conduction analogy, all materials should be metallic with identical real and imaginary indices.

### Field extrema

Another question concerns the ideas of field confinement and enhancement [48-50], which are major challenges in optics. Indeed in optics the harmonic field can spatially oscillate in the stack and specific coatings can be designed to reach huge enhancement of excitation in particular layers. In conduction this would involve a local maximum of temperature within the stack, which is not possible in the absence of sources according to thermodynamics.

This result can be directly recovered from the governing equation (30) in the second Fourier plane, which can be turned into:

$$b\,\hat{u}^{*}\partial^{2}\hat{u}/\partial z^{2} + b\,\alpha^{2}(\sigma,\omega)\,|\hat{u}|^{2} = 0$$

which, after z integration, leads in each medium to:

$$[b\hat{u}^{*}\partial\hat{u}/\partial z]_{z1,z2} - \int_{z} b\,|\partial\hat{u}/\partial z|^{2} + \int_{z} b\,\alpha^{2}(\sigma,\omega)\,|\hat{u}|^{2} = 0 \qquad (54)$$

where the first term within the brackets is continuous. Then the presence of a maximum would allow one to choose these brackets between the $z_1$ location of the extremum, and $z_2$ rejected at infinity where the field is zero. The result would be:

$$\int_{z} b\,|\partial\hat{u}/\partial z|^{2} = \int_{z} b\,\alpha^{2}(\sigma,\omega)\,|\hat{u}|^{2} \qquad (55)$$

which cannot be satisfied either for real or imaginary parts, since $\alpha^2 = j\omega/a - \sigma^2$.

Such a result (no temperature extrema in the absence of sources) does not contradict the fact that thin metallic layers create interferences in optics and allow the field to oscillate with depth location; indeed one can check that optical oscillations vanish when the real and imaginary indices of the metal are identical, which underpins the analogy with conduction. For similar reasons one can show that optical phenomena such as total internal reflection and total absorption [48-50] are not achieved in conduction.

### 5-b Numerical results

For numerical calculation we considered 2 materials with different thermal properties. This allows to keep the analogy with optical multilayers most of which involve 2 materials, that are denoted H (high refractive index $n_H$) and L (low refractive index $n_L$), with thicknesses of the order of a quarter-wavelength ($n_H e_H = \lambda_0/4$, $n_L e_L = \lambda_0/4$). Hence the two thermal materials were chosen with different effusivities $\beta_H$ (high) and $\beta_L$ (low), that are:

$b_H = 418\,W.m^{-1}.K^{-1}$     $a_H = 1.71\,10^{-4}\,m^2/s$ => $\beta_H = 31965\,J.K^{-1}.m^{-2}.s^{-0.5}$     for silver

$b_L = 1.5\,W.m^{-1}.K^{-1}$     $a_L = 7.\,10^{-7}\,m^2/s$ => $\beta_L = 1793\,J.K^{-1}.m^{-2}.s^{-0.5}$     for fused silica

Moreover for each material the thicknesses $e_H$ and $e_L$ were defined with respect to the thermal diffusion lengths at a design pulsation $\omega_0 = 1\,Hz$, that are: $L_H = \sqrt{(2a_H/\omega_0)} = 1.85\,cm$ and $L_L = \sqrt{(2a_L/\omega_0)} = 1.18\,mm$. The thickness values are a quarter of thermal length: $e_H = L_H/4$, $e_B = L_L/4$. Hence the stack under study is a 9 layer coating of design: superstrate/ $e_H e_L e_H e_L e_H e_L e_H e_L e_H$ / substrate, where both superstrate and substrate are quartz.

Then using the optical admittance software we investigated the stack thermal properties versus z location and temporal frequency ($\omega$), at a given spatial frequency $\sigma = 0$. Figure (2) is given for the spectral transmittance $t_{th}(\omega)$ and emphasizes a classical behavior versus increasing frequencies. The frequency range is $[10^{-2}\,Hz, 10\,Hz]$.

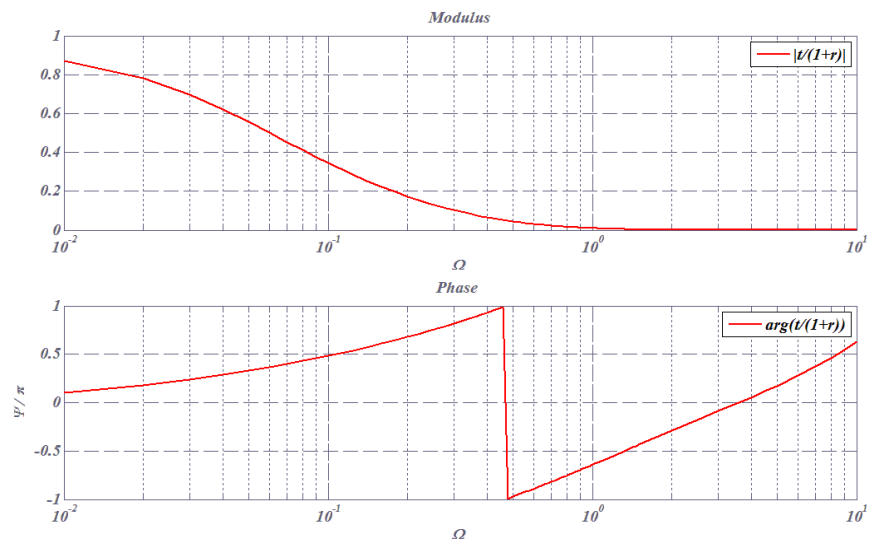

*Figure 2: Spectral thermal transmittance versus pulsation, in modulus and phase. The horizontal units are in Hz. The multilayer is a nine-layer stack designed at pulsation $\omega_0 = 1Hz$.*

The next figures 3-5 (upper and lower panels) are plotted at decreasing working frequencies, that are: $\omega$ = 5Hz, $\omega$ = 1 Hz and $\omega$ = 0.1 Hz. For each frequency the lower curve gives the temperature distribution in modulus within the stack (left vertical scale), with a color scale (right vertical scale) referring to the effusivity of each layer. The admittance diagram can be found in the upper panels, with color circles that enable to connect the admittance position in the complex plane to the altitude within the stack in the lower panels.

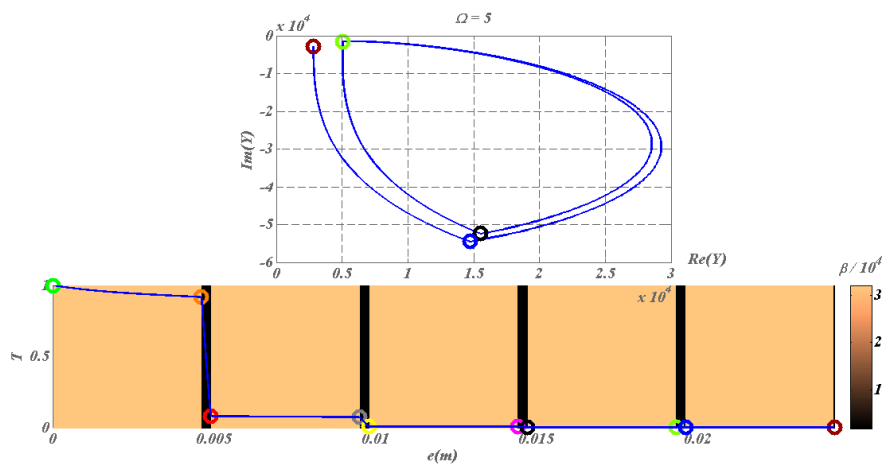

*Figure 3: nine-layer stack at pulsation $\omega = 5Hz$, with the admittance diagram in the complex plane (upper panel) and the distribution of temperature modulus (lower panel)- see text for all scales*

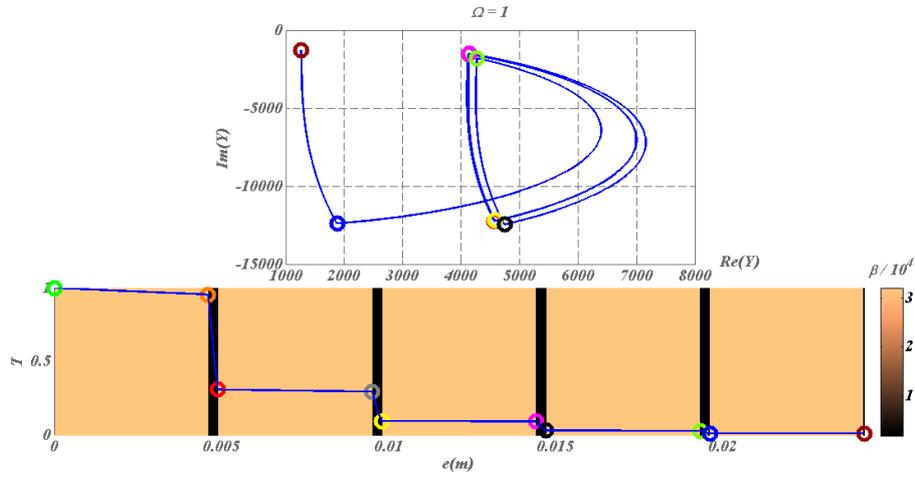

*Figure 4: nine-layer stack at pulsation ω = 1Hz, with the admittance diagram in the complex plane (upper panel) and the distribution of temperature modulus (lower panel)- see text for all scales*

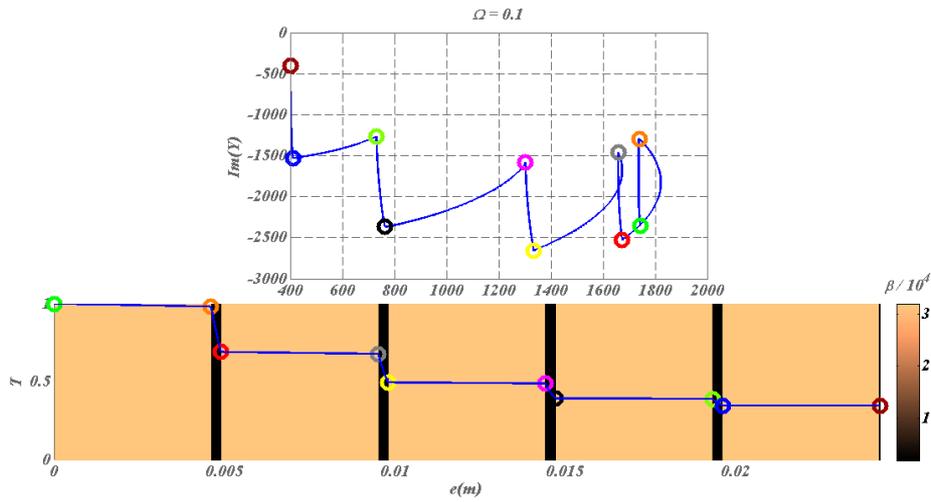

*Figure 5: nine-layer stack at pulsation ω = 0.1Hz, with the admittance diagram in the complex plane (upper panel) and the distribution of temperature modulus (lower panel)- see text for all scales*

It was also necessary to check the energy balance of conduction given in (44). To do that we chose a region free of sources, that is, the stack delimitated by the top interface (in contact with the superstrate) and the bottom interface (in contact with the substrate). Within this domain relation (44) becomes:

$$\hat{S} = 0 \Rightarrow [Y\,\hat{T}]_z = \int_z b\,\alpha^2 \hat{T}\,dz \qquad (57)$$

In figure 6 both real and imaginary parts of these two terms (flux and temperature integral) are plotted and the results show a perfect agreement. We checked that this agreement holds regardless of the spatial frequency σ.

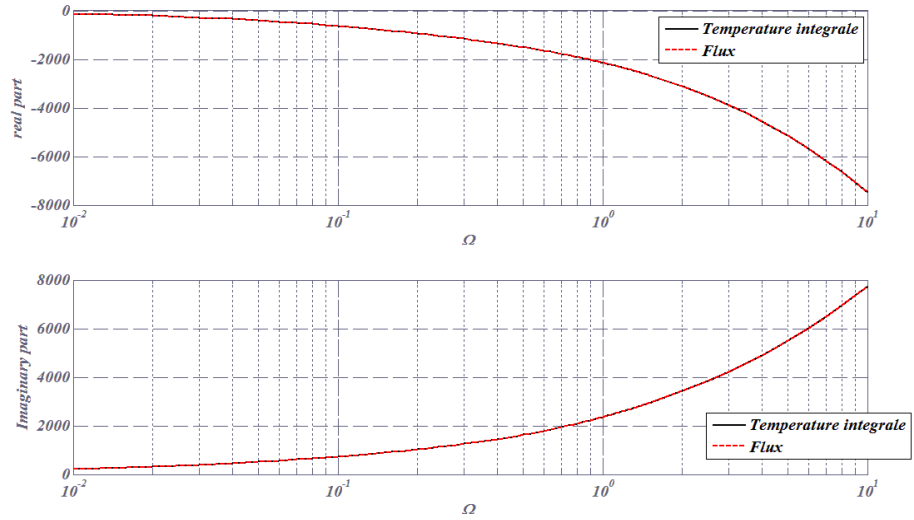

*Figure 6: validation of the complex conduction balance given in relation (57). Solid black curve and red dotted curve are superimposed.*

## 5-c From metallic to transparent optical multilayers

To complete this section figure 7 (a-d) illustrates how the admittance diagrams and the field distribution of metal optics (which we used for conduction) approach those of dielectric multilayers (where the analogy with conduction breaks down) when the imaginary index is decreased. To achieve this correspondence between admittance diagrams in heat conduction and light propagation, we considered a thermal multilayer with conduction number $k = k' + j k''$, and plotted the admittance diagrams at $\Omega = 5Hz$ for decreasing values of $k''$. These values are: $k'' = k'$ (upper left figure), $k'' = 0.3\ k'$ (upper right figure), $k'' = 0.1k'$ (lower left figure) and $k'' = 0$ (lower right figure). The thicknesses are $e_H = L_H/40$ and $e_L = L_L$ at design frequency $\Omega_0 = 1Hz$. These figures confirm that specific admittance circles are progressively recovered when the media become transparent, which is a well-known fact in thin film optics [34]. Also, the field oscillations within the stack are recovered when the imaginary index vanishes, in agreement with (55).

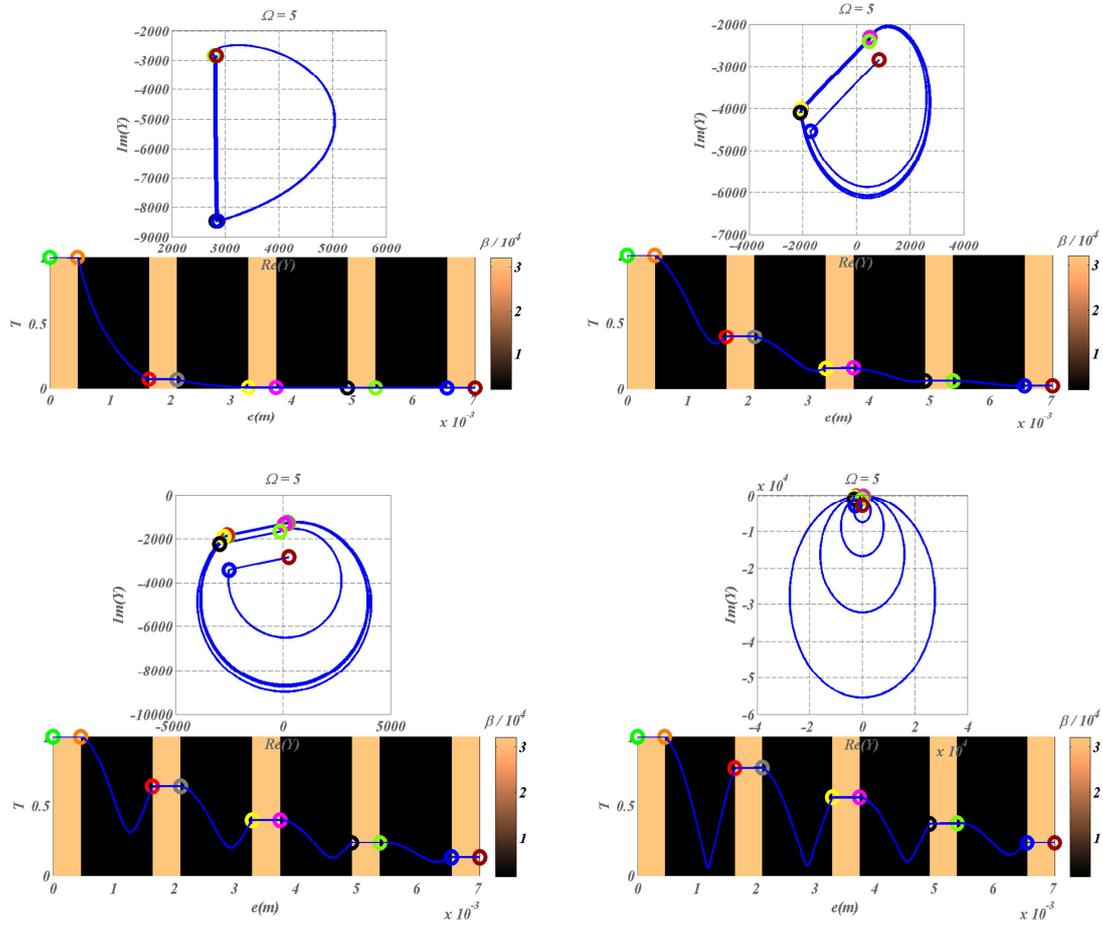

*Figure 7 (a-d): nine-layer metallic stack at pulsation Ω = 5Hz; Modification of admittance diagram (upper panels) and electric field distribution (lower panels) when the imaginary part of the k parameter is decreased from k'' = k' to k'' = 0 (see text for more details). The specific admittance circles of dielectric optics are recovered in the limit of vanishing k''.*

## 6- Micro-cavities

Optical formalisms developed for luminescent micro-cavities also remain valid for heat conduction. Until now multilayers were excited with a source in free space in the superstrate. By opposition with this far field geometry, the same devices are known as micro-cavities when they support the sources in their bulks. The optical (linear) solution [49-50] remains strongly similar whatever the position of the source (far or near field), provided that the field discontinuities are corrected, that is, $\mathbf{z}\wedge\delta\mathbf{E} = \mathbf{z}\wedge\delta\mathbf{H} = \mathbf{0}$ for optical coatings (far field), and $\mathbf{z}\wedge\delta\mathbf{E} = \mathbf{0}$, $\mathbf{z}\wedge\delta\mathbf{H} = \mathbf{J}$ for optical micro-cavities (near field) with $\mathbf{J}$ a surface electric current. Extension of the admittance formalism [49-50] then gives the solution in the second Fourier plane as:

$$\hat{u}_i = \hat{u}'_i = \hat{J}_i/(Y'_i - Y_i) \tag{58}$$

where $\hat{u}_i$ and $\hat{u}'_i$ are the fields on each side of interface i in media i-1 and i respectively, and $\hat{J}_i$ the electric current density at surface i which creates these fields. The admittances characterize each half part of the stack; they are still calculated from (45) but their initial values of effective index are taken in the substrate for Y' and in the superstrate for Y [49-50]. The field sequence (46) can then be used to obtain the field values $\hat{u}_0^{\pm}$ in the extreme media. After summation over all currents the result is:

$$\hat{u}_0^- = \Sigma_i\, C_i^-\hat{J}_i/(Y'_i - Y_i) \qquad \text{and} \qquad \hat{u}_0^+ = \Sigma_i\, C_i^+\hat{J}_i/(Y'_i - Y_i) \tag{59}$$

where $C_i^{\pm}$ are optical factors which are design dependent, and that allow to reach the superstrate ($C_i^-$) or the substrate ($C_i^+$).

Consider now the case of thermal micro-cavities, in the form of planar multilayers supporting thermal sources at theirs interfaces. The conduction process of such devices will again behave like optical cavities in metals, since the discontinuities for temperature and flux as again given in a way similar to optics, that is, for a surface source $S_i$ at interface i:

$$\delta T_i = 0 \text{ and } \delta h_i = S_i \qquad => \qquad T'_i = T_i = \widehat{S}_i/(Y'_i - Y_i) \tag{60}$$

This last relationship is identical to (58) and allows one to address thermal micro-cavities with the same optical softwares, provided that the scalar thermal sources ($S_i$) replace the algebraic electric currents $J_i$. Contrary to the far field geometry, the thermal field now can oscillate within the stack, due to the presence of thermal sources. In the case of bulk sources the field would follow the source at high frequencies. Notice however that optics involves the intensity (the field square), which introduces additional interaction terms ($\hat{J}_i\hat{J}_j^*$) between the currents [49-50]. To conclude this section we notice that the power (optical or thermal) provided by the confined source within the stack depends on the multilayer geometry [49-50].

## 7- Diffraction gratings

Optical diffraction by a periodic surface is well known to spread the incident energy into a discrete series of spatial frequencies given by:

$$\sigma_m = \sigma_0 + q/d \tag{61}$$

with d the grating period, $\sigma_0$ the incident spatial pulsation and q a relative integer. In terms of scattering angles $\theta_p$ in air the result is:

$$\sin\theta_q - \sin i_0 = q\, \lambda/d \tag{62}$$

with $\lambda$ the wavelength and $i_0$ the illumination angle. Several techniques exist to predict the angular efficiency at each diffraction order m, like modal and differential methods, finite elements and boundary integrals…

The same computer codes can again be applied for diffraction of heat conduction, a process which makes temperature only significant at specific frequencies or directions given by (61-62). However unlike for optics one must take account of the attenuation resulting from thermal diffusion, which confines heat diffraction in the vicinity of the grating. Moreover since diffraction does not occur for sub-wavelength gratings, the grating period at normal illumination (excitation) would be greater than one wavelength (d > $\lambda$); such a condition is seemingly trivial in optics but not in conduction, where wavelength and thermal length are similar ($\lambda = 2\pi L$). Indeed the consequence is that the conduction diffraction process will be significant at a distance z of the average surface lower than its period, that is, z < L < d/2π. Finally depending on the thermal length value L, far field effects can be masked or not, which forces diffraction to be emphasized in the near field.

Numerical calculation is given in figures 8-9, at a thermal pulsation of 10 Hz. The grating is a surface with a sinusoidal shape graved at the top of a silver (Ag) substrate, with an amplitude of 1 cm and a period of 10 cm. The superstrate is Air. The heat source results from the absorption of an optical beam of 37.5 cm in size, at normal incidence and 633nm wavelength. The complex refractive index of Ag at this wavelength is chosen as $n_{Ag}$ = 0.134+3.99i. Due to the low depth penetration of optics within the metal, such heat source can be assumed to be located at the grating surface. Moreover the thermal parameters of Ag are 1.71 $10^{-4}m^2/s$ (diffusivity) and 418W/(mK) (conductivity).

The boundary integral formalism for the electromagnetic wave scattering from one-dimensional rough surfaces [52-53] has been adapted to this heat conduction problem. The temperature, which is set continuous across the grating surface, and the normal component of the heat flux, which discontinuity identifies with the heat source, are the unknown functions of a set of coupled boundary integral equations. These equations are numerically solved with the method of moments [51-52] using piecewise-constant basis functions and point-matching testing functions. The discretized grating surface is 3m-long, with a sampling step of 1.5mm.

Figure 8 gives the 2D temperature distribution in the substrate and superstrate. As expected the field shows periodic variations in both media near the center of the beam footprint. However the harmonics which are specific of the grating law cannot be emphasized in the figure, due to the fact that calculation is for the near field, which requires to superimpose all harmonics:

$$\breve{u}(x,z,\omega) = \Sigma_q \ \breve{u}_q(x,z,\omega) \qquad (63)$$

To go further we have plotted in figure 9 the different grating harmonics $\breve{u}_q$ in the superstrate and in the substrate. The results emphasize the validity of the grating law (62) in conduction. Notice that the number of diffraction orders is greater in air (by a factor 3) due to a lower diffusion length.

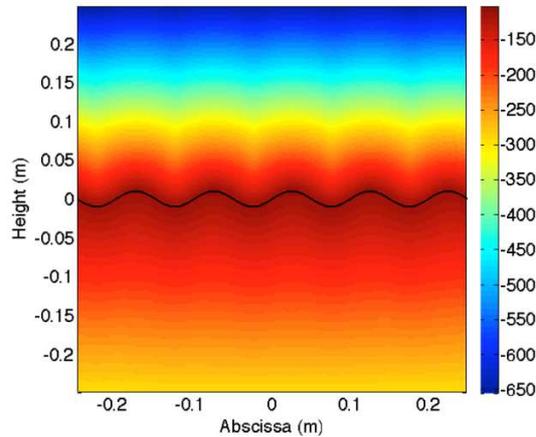

*Figure 8: Diffraction of a conduction process. The full black line is for the sinusoidal surface which separates air (top medium) from silver (bottom medium). The vertical units on the right are for the color levels.*

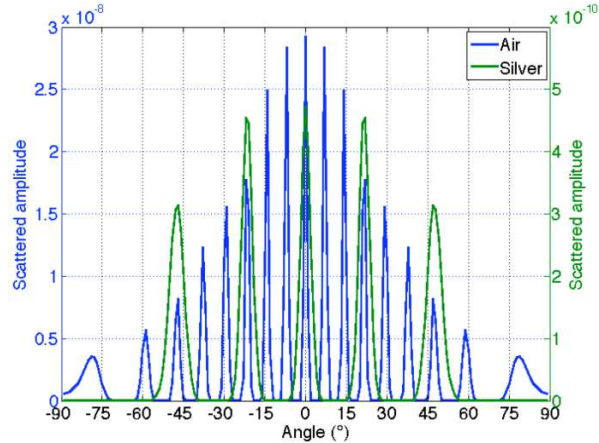

*Figure 9: Spatial harmonics of the temperature (grating law).*

Other diffraction processes can be performed in a similar way, including scattering from inhomogeneities (roughness or bulk) and Mie theory for spherical particles. The diffraction problem for heat conduction could also be addressed within the formalism of microcavities. Indeed a surface thermal source can be created by the deposition of a very thin ($<\lambda/50$) metallic layer on a dielectric material, followed by an etching process with a period d. Under optical illumination, optical absorption in this device will only be concerned with the metallic patterns. When the beam is modulated at frequency $\omega$, the resulting heat source is:

$$S(x,\omega) \approx A(\omega)\ h(x) * \Sigma_q\ \delta(x-qd) = A(\omega)\ \Sigma_n\ h(x-qd) \qquad (64)$$

with $\delta$ the Dirac distribution, h the shape of isolated pattern and A an absorption term. Then Fourier transformation allows one to retrieve the grating law in the form:

$$\hat{S}(\nu,\omega) = A(\omega)\ \Sigma_n\ \hat{h}(q/d)\ \delta(\nu-q/d) \qquad (65)$$

and heat diffraction can be directly calculated following (60).

## 8- Towards a multilayer planar cloak

The last example in the analogy concerns transformation optics, a tool currently used for cloaking in different fields including heat [16-33, 53-54]. Invisibility cloaks usually have circular or spherical geometries, while we keep here the planar geometry of previous sections and consider a single layer which should be "hidden" with the help of additional symmetrical stacks on its left and right sides. In order to design such a planar cloak we apply a geometric transform in the spirit of [54] which maps the disjoint interval $[-e_1,-\eta]$ and $[\eta,e_1]$ (where $\eta<e_1$ is a small positive parameter) onto $[-e_1,-e_2]$ and $[e_2,e_1]$ with $e_2<e_1$. Similarly, the small interval $(-\eta, \eta)$ is mapped onto $(-e_2,e_2)$. When $\eta$ tends to 0, this transform amounts to mapping $[-e_1,e_1]\backslash\{0\}$ onto intervals $[-e_1,-e_2]$ and $(e_2,e_1)$ in the spirit of [30].

While the two transforms make sense in two-dimensional and three-dimensional cases, the former one (Kohn's transform) seems to be more natural in our one-dimensional configuration. With a linear transformation ($x' = \alpha x+\beta$, $y' = y$), we obtain an ideal anisotropic planar cloak surrounding a central (spacer) anisotropic layer. All media (cloak and spacer layer) are anisotropic following the transformation. The conductivity matrix ($b_{ij}$) of the cloak is diagonal with $b_{11} = (e_1-e_2)/(e_1-\eta)$ and $b_{22} = (e_1-\eta)/(e_1-e_2)$, while that ($b'_{ij}$) of the spacer layer is also diagonal with $b'_{11} = e_2/\eta$ and $b'_{22} = \eta/e_2$. In order to achieve these required anisotropic media, we consider an alternation of layers with isotropic conductivities which behave like an effective medium with anisotropic conductivity, see [16,18] for construction of circular counterparts of the planar cloak with homogenization techniques. The result for the cloak is an alternation of isotropic layers of two materials with equal thicknesses and conductivities given at $\eta = 0$ by:

$$b_1 = [1/(e_1-e_2)]\ \{e_1 + [e_2(2e_1-e_2)]^{0.5}\} \quad b_2 = [1/(e_1-e_2)]\ \{e_1 - [e_2(2e_1-e_2)]^{0.5}\} \qquad (66)$$

However although this homogenization process works for the cloak, this is not the case for the spacer layer. Since the spacer conductivity matrix involves two real diagonal terms that are real infinite and zero respectively, the transparency effect with an isotropic spacer layer can only be reached for very large conductivities inside the spacer layer. This prediction is confirmed in figure 10 since the temperature outside the cloak is nearly superimposed to that of the initial homogeneous medium. In this figure the spacer layer conductivity is far greater (several decades) than the other conductivities (that is, $10^5$ higher than that of the initial medium). Complete geometry of the multilayer stack is given in figure 11.

These results confirm that transformation optics still remain valid in the case of heat diffusion. Clearly here, cloaking is limited to some very specific limiting case, due to the high conductivity required within the spacer layer, but the mathematical technique remains the same.

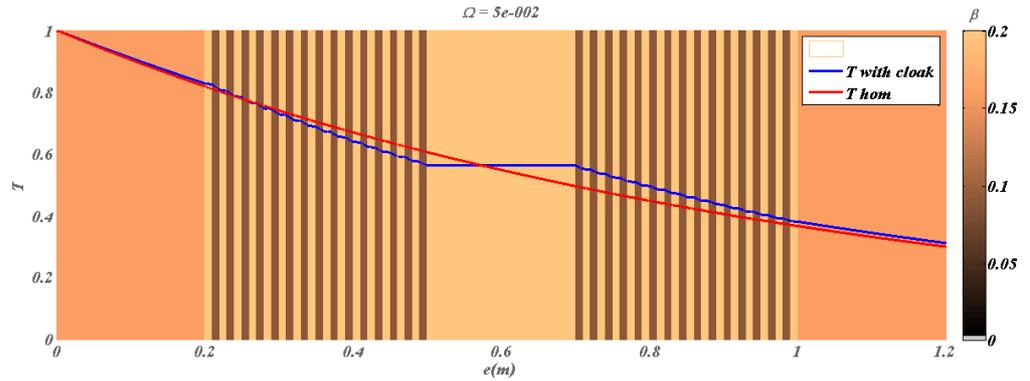

*Figure 10: Field distribution at 0.05Hz inside and outside the multilayer planar cloak surrounding a spacer layer. The field for the homogeneous medium is plotted in red and the field for the highly conducting obstacle surrounded by specially designed stacks (the cloak) is plotted in blue.*

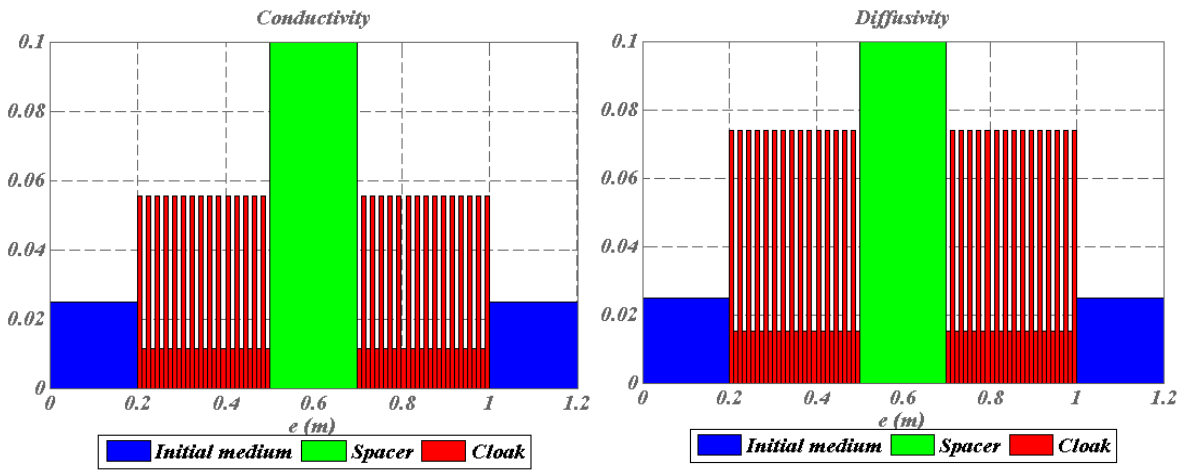

*Figure 11: Stack parameters of figure 10 (conductivities on the left, diffusivities on the right)- The thermal parameters are out of scale for the spacer layer (see text).*

## 9- Static versus harmonic regimes

The matrix formalism (20) developed above works whether the regime is static ($\Omega = 0$) or dynamic ($\Omega \neq 0$). In the static regime the matrix coefficients take a single form and the resulting product matrix for the multilayer becomes:

$$M = \begin{pmatrix} 1 & e_t/b_{eq} \\ 0 & 1 \end{pmatrix} \quad (67)$$

with $e_t$ the total thickness and $b_{eq}$ the equivalent conductivity:

$$e_t/b_{eq} = \Sigma_k e_k/b_k \qquad (68)$$

which is a well-known result. However there is a key difference with the harmonic regime which concerns the admittances. Indeed at non zero frequencies the admittances only depend on the multilayer design (geometry, optical or thermal parameters), while in the static regime they also depend on the extreme fields in the form:

$$Y_s = (b_{eq}/e_t)(T_0-T_s)/T_s \quad \text{and} \quad Y_0 = (b_{eq}/e_t)(T_0-T_s)/T_0 \qquad (69)$$

This point recalls why the frontier temperatures (or fluxes) must be forced in the static regime, which is not required in the harmonic regime. This is related to the fact that the 1D static solution in the extreme media free of sources involves 2 parameters ($T = \alpha z + \beta$), while the harmonic solution requires 1 parameter ($T = T_s \exp(j\alpha z)$); in other words, with the harmonic regime the derivative of T is proportional to T, which is not the case for the static solution.

The analogy with optics still holds in the static regime since electrostatics also involves a potential difference (rather than a potential), which is classical. However in the harmonic regime (electromagnetism) there is no supplementary condition at the domain frontiers, so that the multilayer problem is entirely solved by the sequence relationship (45) of the admittances, which allows one to retrieve the top admittance $Y_0(m_s)$ from the substrate one. However in some situations like modal optics (guided waves) another condition must be fulfilled to force the field to only merge both in the superstrate and substrate, that is : $Y_0(m_s) = -m_0$. The result is a discretization of (45) which gives the guided modes as the poles of the reflection factor. A similar discretization would be obtained in the harmonic regime when temperature (or flux) is forced in both extreme media.

## 10-    Conclusion

We have explored an analogy between optical propagation and heat conduction in isotropic, homogeneous and linear media. Both fields (scalar electric field and temperature) were shown to follow the same harmonic equation in the second Fourier plane, provided that the optical wave number $k_{opt} = \omega\sqrt{[\check{\varepsilon}(\omega)\check{\mu}(\omega)]}$ is replaced by the heat conduction number $k_{th} = (1+j)\sqrt{[\omega/2a]}$. The dispersion law of this wave number has the memory of the time derivation order of the governing spatio-temporal equation. Moreover, the analytic expression of the fields, conduction and wave numbers show that thermal diffusion is strongly analogous to optical propagation in metallic media; however the optical refractive index of these artificial metallic media should have identical real and imaginary parts, a specific property where applications of metal optics can be reduced.

The case of planar multilayers was addressed and the optical admittance formalism was shown to be directly applicable to thermal conduction. Within this framework the temperature and algebraic heat flux played the role of electric and magnetic field tangential components, respectively. The only noticeable difference in the admittance formalism is the required modification of the optical effective index in order to take into account the heat conduction number ($m_{th} = -j\alpha b$ replaces $m_{opt} = n\alpha/k$). Numerical calculations were presented to validate all results, including the optical and thermal energy balances.

The analogy between metal optics and conduction was then extended to micro-cavities and diffraction gratings, and to transformation optics. All results show that most softwares developed for metal optics can be directly used for conduction, whether they address interference filters or micro-cavities, diffraction gratings or photonic crystals, meta-materials or transformation optics. We finally note that Green's functions for reflected and thermal fields at planar interfaces were computed in [53] and they are reminiscent of what researchers compute in optics. However one must keep in mind that these analogies only hold for specific metals (real index = imaginary index). We hope our work will foster theoretical and experimental efforts in thermal metamaterials.

## Acknowledgments
This work was supported by the ANR (Agence Nationale de la Recherche) and DGA (Direction Générale de l'Armement).